\documentstyle[12pt,aaspp]{article}
\newcommand{\fulljustify}{\relax}
\fulljustify     % This must appear in the abstract as well, so to toggle
\newcommand{\beq}{\begin{equation}}              % end equation
\newcommand{\eeq}{\end{equation}}             % end equation
\newcommand{\beqa}{\begin{eqnarray}}              % end equation array
\newcommand{\eeqa}{\end{eqnarray}}             % end equation array

\def\j21{J_{21}}

\def\lya{Ly$\alpha$~}

\def\myref{\bibitem{dummy}\vspace{-2pt}}
\let\gsim=\gtrsim
\let\lsim=\lesssim
\begin{document}
\title{Microwave Background Anisotropies Due to the \\
Kinematic Sunyaev-Zel'dovich Effect of the \lya Forest} 
\author{Abraham Loeb}
\affil{Astronomy Department, Harvard University\\ 
60 Garden St., Cambridge MA 02138}

\begin{abstract}
\fulljustify
The \lya absorption systems observed in the spectra of QSOs are likely to
possess bulk peculiar velocities. The free electrons in these systems
scatter the microwave background and distort its spectrum through the
kinematic Sunyaev-Zel'dovich effect.  I calculate the temperature
fluctuations of the microwave sky due to variations in the number
of \lya systems along different lines-of-sight throughout the universe.
The known population of absorbers out to $z\approx 5$ introduces
anisotropies on angular scales $\lsim 1^\prime$ with an rms amplitude of
order $\Delta T/T\approx 10^{-6} (\Omega_{\rm Ly\alpha}/0.05)\langle
v_{400}^2\rangle^{1/2}$, where $\Omega_{\rm Ly\alpha}$ is the cosmological
density parameter of ionized gas in \lya absorption systems, and $\langle
v_{400}^2\rangle^{1/2}$ is the rms line-of-sight peculiar velocity of these
systems at $z\sim 3$ in units of $400~{\rm km~s^{-1}}$.  Detection of this
signal will provide valuable information about the cosmic velocity field
and the gas content of \lya absorption systems at high redshifts.
 
\bigskip
\noindent
\end{abstract}

\bigskip
\bigskip
\centerline{Submitted to {\it The Astrophysical Journal Letters}, April 1996}

\clearpage
\section{Introduction}

Sunyaev and Zel'dovich showed that Compton scattering off free electrons
would distort the spectrum of the cosmic background radiation (CBR) either
if the electrons have a different temperature than the radiation
or if they possess a bulk velocity relative to the cosmic frame (see
Sunyaev \& Zel'dovich 1980 for a review). By now, the thermal effect has
been seen in many clusters of galaxies (Birkinshaw 1993; Rephaeli 1995),
and observations to measure the kinematic effect are underway (see
discussion in Haehnelt \& Tegmark 1995).  The integrated level of
distortion introduced to the CBR spectrum by intergalactic gas is currently
unknown, although it may be considerable in popular cold dark matter
cosmologies (Colafrancesco et al. 1994, 1995; Persi et al. 1995).  The
perfect blackbody spectrum detected by the {\it COBE} satellite sets a
tight limit on the total Comptonization parameter of the universe,
$y\leq1.5\times10^{-5}$ (Fixsen et al. 1996).  Although this limit can be
used to rule-out exotic cosmological scenarios (Levin, Freese,
\& Spergel 1992; Wright et al.  1994), it still allows for a significant
contribution to the observed CBR anisotropies by the Sunyaev-Zel'dovich
(SZ) effect. As future satellite experiments are currently being designed
with the hope to map the microwave sky on angular scales $\gsim 10^\prime$
(Janssen et al. 1996), it is important to characterize the unavoidable
contribution from intergalactic scattering to the expected anisotropies.

The intergalactic medium at redshifts $z\lsim 5$ is highly ionized and
inhomogeneous.  Direct evidence for condensations of gas comes from the
absorption lines observed in QSO spectra (see reviews in Blades, Turnshek,
\& Norman 1988).  Close QSO pairs and multiple images of gravitationally
lensed QSOs indicate that the transverse extent of these absorption
systems is $\ell_{\perp}\sim0.1$--$1{\rm Mpc}$ (Bechtold et al. 1994;
Dinshaw et al.  1994, 1995; Fang et al. 1995; Smette et al. 1995).  The
neutral hydrogen density in these systems is thought to be controlled
through photoionization equilibrium by an ionizing radiation background.
The existence of this background radiation has been established by the
``proximity effect'', with the background flux estimated from the size of
the ionization region surrounding bright QSOs
%--inside which the QSO flux dominates over the background flux 
(e.g., Bajtlik, Duncan, \& Ostriker
1988; Lu, Wolfe, \& Turnshek 1991; Bechtold 1994). Hydrodynamic simulations
of popular cosmological models tend to recover the essential properties of
the observed absorption systems and imply that these systems are a generic
consequence of early structure formation in the universe (Cen et al. 1994;
Zhang et al. 1995; Hernquist et al. 1996; Miralda-Escud\'e et al. 1995).

Like other objects in the universe, the \lya absorption systems are
expected to develop bulk peculiar velocities in response to the
gravitational field in their vicinity. Thus, as the free electrons in an
absorption system scatter the microwave background, they would distort its
spectrum through the kinematic Sunyaev-Zel'dovich effect. For a single
absorption systems with a line-of-sight peculiar velocity of $\vert v\vert\gsim
100~{\rm km~ s^{-1}}$ and a characteristic temperature in the range $T\sim
10^{4-5}~{\rm K}$, the kinematic SZ effect is much larger than the thermal
SZ effect since $\vert v/c\vert\gsim 10\times (2kT/m_ec^2)$, where $m_e$ is the
electron rest mass.  This is the opposite situation from that in clusters of
galaxies. The Rayleigh-Jeans distortion is then  $(\Delta
T/T)\approx -\tau v/c$, where $\tau$ is the optical depth
of the \lya system to Thomson scattering. The contributions from positive and
negative velocities of individual systems would tend to cancel out on
average, and the integrated CBR spectrum over the whole sky would therefore
show little evidence for this distortion.  However, in small regions of the
sky, of order the angular scale occupied by a single \lya absorption system
($\lsim 1^{\prime}$), there could be large rms amplitude of fluctuations
due to incomplete cancelations between a small number of absorption
systems. In this {\it Letter}, I calculate these fluctuations and show that
their detection could provide important new information about \lya
absorption systems at high redshifts.

\section{Calculation}
%The Kinematic Sunyaev-Zel'dovich Effect of the \lya Forest}

Our discussion is based on the assumption that the intergalactic gas at
high-redshifts is in photoionization equilibrium with a UV radiation
background.  The time required to establish photoionization equilibrium,
$\sim 10^4 J_{21}^{-1}~{\rm yr}$, is much shorter than the Hubble time, for
the Lyman-limit intensity inferred from the proximity effect (Bechtold
1994), $J_\nu= J_{21}\times 10^{-21}{\rm
erg~cm^{-2}~s^{-1}~Hz^{-1}~sr^{-1}}$. We focus our attention on systems
with column densities below the Lyman limit, $N\leq N_{\rm max}=
10^{17}~{\rm cm^{-2}}$, because thicker systems are opaque to
photoionizing photons and are therefore composed primarily of neutral
hydrogen that does not interact with the microwave background.
%The contribution to the microwave
%anisotropies from the outer, mean free path thick, ionized surfaces of
%the high column-density systems is small because of the low abundance and
%the small angular size of these systems.

Let us first find the rms amplitude of fluctuations in the optical depth of
the universe due to the clumpiness of the intergalactic gas.  Most of this
gas is known to be ionized and its neutral hydrogen component serves as a
tracer of the fluctuations in the electron density.  At photoionization
equilibrium, the clumpiness in the electron density is directly
proportional to the local neutral hydrogen density, $n_e^2\propto n_{\rm
H}$. Thus, overdense regions show up as absorption lines in QSO spectra and
their statistical properties can be inferred from observations.  We define
$n_e$ to be the mean electron density within an absorption system of
thickness $\ell$ and HI column density $N$.  At photoionization
equilibrium,
\beq\label{phion}
n_e^2 \ell = \eta N
\eeq
where $\eta\approx 30 J_{21} (T/5\times 10^{4}~{\rm K})^{0.7}~{\rm
cm^{-3}}$ (Spitzer 1978; Peebles 1993). For the redshift interval of
interest $2\lsim z\lsim 5$ we assume that $J_{21}$ and therefore $\eta$ are
constant in time (see discussions in Bajtlik et al.\ 1988 and Bechtold 1994).
The optical depth of an absorption system to Thomson scattering can then be
obtained from equation\ (\ref{phion}),
\beq\label{optdepth}
\tau(N)=\sigma_T n_e {\bar{\ell}}= \sigma_T (\eta N {\bar{\ell}})^{1/2} ,
\eeq
where $\sigma_T$ is the Thomson cross-section, and ${\bar{\ell}}(N)$ is the
ensemble averaged thickness of all absorption systems with column density
$N$. For simplicity, we parametrize ${\bar{\ell}}=B N^{-\alpha}$.  While
low-column density systems with $N\sim 10^{14}~{\rm cm^{-2}}$ may have a
line-of-sight thickness ${\bar{\ell}} \lsim 10^{2}~{\rm kpc}$ 
(Rauch \& Haehnelt 1995; Bechtold et al. 1994;
Dinshaw et al. 1994, 1995; Fang et al. 1995; Smette et al. 1995), damped 
\lya absobers with $N\sim 10^{21}~{\rm cm^{-2}}$
are probably proto-galactic disks with a characteristic thickness
${\bar{\ell}}\sim 10^{0-1}~{\rm kpc}$.  This modest variation in thickness
over many decades in column density implies a relatively small value of
$\alpha\sim 0.2$. As shown later, the value of $B$ does not enter into our
final result.

The probability of intersecting a \lya absorption system with an HI column
density between $N$ and $N+dN$ in a redshift interval $dz$ around $z\sim 3$ is
well fitted by a power-law (Carswell et al. 1984; Tytler 1987),
\beq\label{powerlaw}
\left({dP \over dN dz}\right)_{z=3} = A N^{-\beta},
\eeq
where $A= 10^{9.1\pm0.3}~{\rm cm^{[2\beta+2]}}$ and $\beta= 1.51\pm0.02$
(Sargent, Steidel, \& Boksenberg 1989).  Most of the fluctuations in the
optical depth of the universe are contributed by systems close to the
Lyman-limit. For these systems we extrapolate equation\ (\ref{powerlaw}) to 
other redshifts with the scaling,
\beq\label{theredshiftdep}
{dP\over dN dz}\propto (1+z)^{\gamma},
\eeq
where $\gamma=1.5\pm0.4$ (Stengler-Larrea et al. 1995; see also
Storrie-Lombardi et al. 1994).  Since the systems of interest are separated
by large redshift intervals along a given line of sight (e.g., there is
typically one Lyman-limit system per unit redshift), we can safely ignore
any correlations between them (see also Sargent et al. 1980; Webb \&
Barcons 1991) and assume that they are randomly distributed in redshift
and column density according to the probability distribution in equations\
(\ref{powerlaw}) and\ (\ref{theredshiftdep}).

We can now calculate the rms level of fluctuations in the optical depth out
to $z=5$ due to the finite number of absorption systems along random lines
of sight. In each infinitesimal $(\Delta N,\Delta
z)$ bin there is either one or no \lya absorption system.  The probability
of having a system with an optical depth $\tau(N)$, is $p= (dP/dN dz)\Delta N
\Delta z\ll1$. Therefore, the variance of $\tau$ 
in each bin is $p\tau^2-(p\tau)^2\approx
p\tau^2$.  The integrated variance is then obtained by summing the
independent contributions of all infinitesimal bins,
\beq\label{vartau}
\langle\Delta\tau^2\rangle=\int_0^5 dz \int_{N_{\rm min}}^{N_{\rm max}}
dN {dP\over dN dz}\times \tau^2=\left[{4.4 \eta A B \sigma_T^2\over
2-\beta-\alpha}\right] 
\left(N_{\rm max}^{2-\beta-\alpha} -N_{\rm min}^{2-\beta-\alpha}\right),
\eeq
where the second equality follows from equations\
(\ref{optdepth})--(\ref{theredshiftdep}) with $\gamma=1.5$. As expected,
the total rms amplitude of fluctuations grows in proportion to the square-root
of the number of \lya systems, namely
$\langle\Delta\tau^2\rangle^{1/2}\propto A^{1/2}$.  The constant
coefficient in square brackets depends on a variety of uncertain
parameters, such as $J_{21}$, $T$, or $B$. We therefore prefer to trade
these parameters in favor of a single quantity, the density parameter of
ionized gas, whose value can be related to Big-Bang nucleosynthesis.  We
achieve this conversion by calculating the derivative of the mean optical
depth of the universe from equations\ (\ref{optdepth}) and\
(\ref{theredshiftdep}),
\beq\label{meantau}
\left({d\langle\tau\rangle\over dz}\right)_{z=3}=
\int_{N_{\rm min}}^{N_{\rm max}}
dN \left({dP\over dN dz}\right)_{z=3}\times \tau =
\left[{A ({\eta B})^{1/2}\sigma_T\over 3/2-\beta-\alpha/2}\right] 
\left(N_{\rm max}^{3/2-\beta-\alpha/2}-
N_{\rm min}^{3/2-\beta-\alpha/2} \right).
\eeq
Since the mean optical depth is invariant to clumpiness for an
optically-thin medium, we can also express $d\langle\tau\rangle/dz$
in terms of the density parameter of ionized gas in absorption
systems below the Lyman-limit, $\Omega_{\rm Ly\alpha}$, 
\beq\label{newmeantau}
\left({d\langle\tau\rangle\over dz}\right)_{z=3}=
-\left(\langle n_e\rangle \sigma_T 
{c dt\over dz}\right)_{z=3}  = 5.5\times 10^{-3} 
\left[{F(\Omega,\Omega_{\Lambda},3)\over F(0.3,0.7,3)}\right]
\left({\Omega_{\rm Ly\alpha}h_{50}\over 0.05}\right),
\eeq
where $h_{50}$ is the Hubble constant in units of $50~{\rm
km~s^{-1}~Mpc^{-1}}$, $\Omega$ and $\Omega_{\Lambda}$ are the density
parameters of matter and the cosmological constant respectively, and
$F(\Omega, \Omega_{\Lambda},z)\equiv
(1+z)^2/[\Omega(1+z)^3+(1-\Omega-
\Omega_{\Lambda})(1+z)^2+\Omega_{\Lambda}]^{1/2}$.

The density of baryons predicted by Big-Bang nucleosynthesis could either
be in the form of ionized gas, neutral gas, or compact objects such as
stars.  The neutral hydrogen can be weighted by the frequency of absorption
systems in the spectra of QSOs; this yields $\Omega_{\rm HI}\approx
(1-4)\times10^{-3}$ at a redshift $z\sim 3$, a value which is comparable to the
mass density of luminous stars in the local universe (Lanzetta, Wolfe,
\& Turnshek 1995).  The metalicity of the densest absorption systems which 
are thought to be the progenitors of present-day disk galaxies (the
so-called damped \lya systems), is only a tenth of the solar value
(Lanzetta et al. 1995; Pettini, Lipman, \& Hunstead 1995; Sembach et al.
1995), implying that star formation is only at its infancy at redshifts
$z\sim 2$--$5$.  If we therefore assume that the stellar mass fraction is
small with a density parameter $\Omega_\star\ll\Omega_{\rm HI}$, then
most of the baryons in the universe may be in the form of ionized gas at
$z\sim 2$--$5$.  Based on the baryonic density parameter predicted by
standard Big-Bang nucleosynthesis (Walker et al. 1991), $\Omega_b\approx
0.05 h_{50}^{-2}$, it is then plausible to substitute $\Omega_{\rm
Ly\alpha}\sim 0.05$ in equation\ (\ref{newmeantau}). Indeed, hydrodynamic
simulations indicate that more than $85\%$ of the baryons reside in
absorption systems below the Lyman-limit (see Fig. 23
in Miralda-Escud\'e et al. 1995).

By multiplying the square--root of equation\ (\ref{vartau}) with the ratio
of equations\ (\ref{newmeantau}) and\ (\ref{meantau}) we get
\beq\label{finvartau}
\langle\Delta\tau^2\rangle^{1/2}= 1.2\times 10^{-2}
\left[{F(\Omega,\Omega_\Lambda,3)\over 
F(0.3,0.7,3)}\right]
\left({\Omega_{\rm Ly\alpha}h_{50}\over 0.05}\right)
{(-3/2+\beta+\alpha/2)\over [A(2-\beta-\alpha)]^{1/2}}
{[N_{\rm max}^{(2-\beta-\alpha)}-N_{\rm min}^{(2-\beta-\alpha)}]^{1/2}
\over [N_{\rm min}^{(3/2-\beta-\alpha/2)}-N_{\rm max}^{(3/2-
\beta-\alpha/2)}]} .
\eeq
For $A=10^{9.1}~{\rm cm^5}$, $\beta=1.5$, $\Omega_{\rm
Ly\alpha}h_{50}=0.05$, $\Omega=0.3$, $\Omega_\Lambda=0.7$, $\alpha=0.2$,
$N_{\rm max}=10^{17}~{\rm cm^{-2}}$, and $N_{\rm min}=10^{13}~{\rm
cm^{-2}}$, we get $\langle\Delta\tau^2\rangle^{1/2}=7\times10^{-4}$. This
corresponds to an rms amplitude of fluctuations of order $3\%$ in the mean
optical depth of the universe out to $z=5$. This result has a weak
dependence on uncertainties in the values of most of the above parameters
(to a first approximation, $\langle\Delta\tau^2\rangle^{1/2}\propto
A^{-1/2} N_{\rm min}^{0.1}N_{\rm max}^{0.15}$).

We finally wish to calculate the resulting fluctuations in the microwave
background temperature due to the kinematic Sunyaev-Zel'dovich effect of
the \lya forest. For this purpose, we assume that the optical depth
$\tau$ and the peculiar velocity $v$ of an absorption system are
uncorrelated and then repeat the calculation that led to equation\
(\ref{vartau}). Note that we define $v$ to be the net bulk velocity of an
absorption system, and so we average over internal velocities within the
system.  The resulting rms amplitude of fluctuations in the microwave
temperature is,
\beq\label{deltat}
\left({\Delta T\over T}\right)={\langle
\Delta\tau^2\rangle^{1/2}}{\langle ({v/c})^2\rangle^{1/2}}
\approx 10^{-6} \left({\Omega_{\rm Ly\alpha}\over 0.05}\right)
\langle\left({v\over 400~{\rm km~s^{-1}}}\right)^2\rangle^{1/2},
\eeq
where $\langle v^2\rangle$ is the dispersion in the line-of-sight peculiar
velocities of absorption systems, and the second equality was obtained for
the set of parameter values chosen after equation\ (\ref{finvartau}).  The
dispersion $\langle v^2\rangle$ is averaged over redshift with a weight
factor of $(1+z)^\gamma$.  The anisotropy signal expressed in equation\
(\ref{deltat}) varies on the angular size of a \lya absorption system at
$z\sim 5$, $\theta_{\rm Ly\alpha} \sim 1^{\prime}
\times ({\ell_{\perp}/500 h_{50}^{-1}~{\rm
kpc}})$, and declines on larger scales roughly as $\propto
(\theta/\theta_{\rm Ly\alpha})^{-1}$.

In the local universe, the peculiar velocity field of galaxies shows a
characteristic one-dimensional dispersion of order $\sim 500~{\rm
km~s^{-1}}$ (Strauss, Cen, \& Ostriker 1993) and a coherence length of
order a few tens of Mpc (G\'orski et al.  1989; Strauss \&
Willick 1995).  Both linear theory (yielding $v\propto (1+z)^{-0.5}$ for
$\Omega=1$ and a more moderate decline for $\Omega<1$) and nonlinear
dynamics predict that the velocity dispersion would scale down only by a
modest factor of order $2$ for $z\sim3$. It is therefore plausible to
consider $\langle v^2\rangle^{1/2}\sim 300~{\rm km~s^{-1}}$ in equation\
(\ref{deltat}). The redshift evolution of the \lya absorber population
conspires to almost cancel out the expected decline in the rms peculiar
velocity with redshift; for $\Omega=1$ equations\
(\ref{vartau})--(\ref{newmeantau}) yield
$\langle\Delta\tau^2\rangle^{1/2}\propto
(1+z)^{1-\gamma/2}=(1+z)^{0.25}$. One should remember, however, that in
difference from galaxies, pressure forces can influence the
peculiar velocity field of \lya absorption systems, and so naive scalings
based on the theory of collisionless gravitational instability may not
provide accurate predictions for these systems.  Since the SZ fluctuations
calculated above are dominated by rare high column--density systems which
are separated by a path length much longer than the coherence length of the
peculiar velocity field, we are justified in ignoring any velocity
correlations between neighboring systems on angular scales $\sim
\theta_{\rm Ly\alpha}$.

\clearpage
\section{Discussion}

Equation\ (\ref{deltat}) summarizes the expected temperature fluctuations in
the microwave sky due to the kinematic Sunyaev-Zel'dovich effect of discrete
\lya absorption systems. This result is based on the observed statistical
properties of the \lya forest in the spectra of QSOs (cf. Eq.\
(\ref{finvartau})).  The predicted anisotropy signal appears on angular
scales $\sim 0.1$--$1^\prime$ and decreases roughly as $\propto
\theta^{-1}$ on larger scales.  Existing VLA observations place an upper
limit of $(\Delta T/T)<2\times 10^{-5}$ on an angular scale of
$1.3^{\prime}$ (Fomalont et al. 1993), which is consistent with the
prediction of equation\ (\ref{deltat}).  The predicted effect dominates
over the anisotropies produced by Bremsstrahlung emission from the same
\lya absorption systems (Loeb 1996) for
wavelengths $\lambda \lsim 3~{\rm cm}$.

The magnitude of the expected signal from the \lya forest is comparable to
the Ostriker--Vishniac effect, which evaluates the anisotropies due to bulk
velocities during early reionization using second-order perturbation theory
(Ostriker \& Vishniac 1986; Vishniac 1987; Hu \& White 1995).  However, it
is important to emphasize that the predictions of this {\it Letter} rely on
the observed population of \lya absorption systems with a minimal level of
speculation about the full reionization history of the universe. While the
{\it thermal} SZ effect from clusters of galaxies may also add fluctuations
of comparable magnitude on the same angular scale (Colafrancesco et al.
1994, 1995), it is possible to separate it from the {\it kinematic} SZ
effect of \lya systems either by spectral measurements in several microwave
bands, or by a cross-correlation analysis with the fluctuations of the
X-ray sky. The dominant contamination to the signal from \lya systems comes
from radio sources (see Figs. 9--11 in Tegmark \& Efstathiou 1996).
Although their contribution to the anisotropies on arcminute scales is
large, radio sources could in principle be removed through high-resolution
multiple-frequency observations (Fomalont et al. 1993; Windhorst et al.
1993).

If the anisotropies caused by \lya absorption systems are detected, they would
provide valuable information about the cosmic velocity field and the gas
content of these systems at high redshifts.  The two-dimensional
information imprinted on the microwave sky could in principle
be extended into redshift space, through a cross-correlation analysis
with absorption spectra of QSOs in the same region of the sky.
%(after the appropriate subtraction of the radio loud QSOs 
%from the microwave signal).
When extended surveys of QSOs, such as the forthcoming Sloan Digital Sky
Survey (Gunn \& Weinberg 1995), become available, it will be possible
to perform such a correlation study over large areas of the sky.

\acknowledgements
I thank Daniel Eisenstein, Tsafrir Kolatt, and Uros Seljak
for useful discussions. This work was supported in
part by the NASA ATP grant NAG5-3085.

\end{document}